\documentclass[aps,pra,twocolumn,showpacs,superscriptaddress,groupedaddress]{revtex4-1}
\usepackage{graphicx}  
\usepackage{dcolumn}   
\usepackage{bm}        
\usepackage{amssymb}   
\usepackage{amsmath}
\usepackage{amsfonts}
\usepackage[utf8]{inputenc}
\usepackage{amssymb}
\usepackage[caption=false]{subfig}
\usepackage{changes}
\usepackage{hyperref}

\renewcommand{\vec}[1]{\mathbf{#1}}

\begin{document}


\title{Photon-induced Self Trapping and Entanglement of a Bosonic Josephson Junction Inside an Optical Resonator}

\author{P. Rosson$^{1}$, G. Mazzarella$^{1}$, G. Szirmai$^{2}$, L. Salasnich$^{1,3}$}
\affiliation{$^{1}$Dipartimento di Fisica e Astronomia ``Galileo Galilei''
and CNISM, Universit\`a di Padova, Via Marzolo 8, I-35131 Padova, Italy
\\
$^{2}$Institute for Solid State Physics and Optics -
Wigner Research Centre for Physics,
Hungarian Academy of Sciences, H-1525 Budapest P.O. Box 49, Hungary
\\
$^{3}$CNR-INO, via Nello Carrara, 1 - 50019 Sesto Fiorentino, Italy}

\date{\today}

\begin{abstract}
We study the influence of photons on the dynamics and the ground state of the atoms in a Bosonic Josephson junction inside an optical resonator. The system is engineered in such a way that the atomic tunneling can be tuned by changing the number of photons in the cavity. In this setup the cavity photons are a new means of control, which can be utilized both in inducing self-trapping solutions and in driving the crossover of the ground state from an atomic coherent state to a Schr\"odinger's cat state. This is achieved, for suitable setup configurations, with interatomic interactions weaker than those required in the absence of cavity. This is corroborated by the study of the entanglement entropy. In the presence of a laser, this quantum indicator attains its maximum value (which marks the formation of the cat-like state and, at a semiclassical level, the onset of self-trapping) for attractions smaller than those of the bare junction.


\end{abstract}

\maketitle

\section{Introduction}

A Bose Einstein condensate confined in a double-well potential mimics in many ways the coherent dynamics of a superconducting Josephson junction \cite{josephson,smerzi}. Therefore, it is often called a Bosonic Josephson junction (BJJ), whose coherent dynamics can be described by the nonlinear equations of a non-rigid pendulum \cite{smerzi}. Furthermore, the investigation of the BJJ dynamics allows us to study the formation of macroscopic coherent states \cite{smerzi,stringa} and macroscopic Schr\"{o}dinger's cat states \cite{cirac,diaz,diaz2,huang,brand,carr2010dynamical,garcia2012macroscopic}.
When the condensate is inside an optical resonator, the photon and atomic degrees of freedom are no longer independent \cite{ritch2013cold}: there is a mutual back-action between the cavity and BJJ dynamics. The photon field acts as an optical potential on the atoms; at the same time, the atomic density serves as a refractive medium for the cavity photons. Note that the idea underlying this scheme is closely related to the models discussed, within the optical lattices context, in \cite{maschler2007,maschler2008,caballero2015}.
In an earlier paper it was shown, how the constant photon field of a laser - tightly focused to the center of the junction barrier - influences the tunneling properties by effectively raising or lowering the barrier, or ultimately allowing for a localized state inside the central well \cite{laserdip}. Later it was also investigated, how the dynamical nature of a cavity field - focused similarly to the center of the barrier - alters the BJJ dynamics in the semiclassical level \cite{articolo}.
The purpose of this paper is to study how the photons influence the exact quantum dynamics and the ground state of the junction, when the parameters of the system are chosen in such a way that the cavity field can be considered as fixed, i.e. in this bad cavity case the photon field is practically a laser field. The cavity photons alter the equations that describe the BJJ dynamics \cite{articolo}, and we investigate how they can induce self-trapping solutions.
The study of the transition from the Josephson regime to the self-trapping regime in the semiclassical dynamics is complemented by the analysis of the ground state. The crossover from the atomic coherent state to the Schr\"{o}dinger's cat state, can in fact be regarded as the quantum counterpart to the semiclassical examination of the system.
The ground state of the system is analyzed using three quantum indicators: the Fisher information, the coherence visibility and the entanglement entropy. This last estimator proves to play a central role in understanding when the transition takes place.

\section{The model}
\begin{figure}[tb!]
\centering
\includegraphics{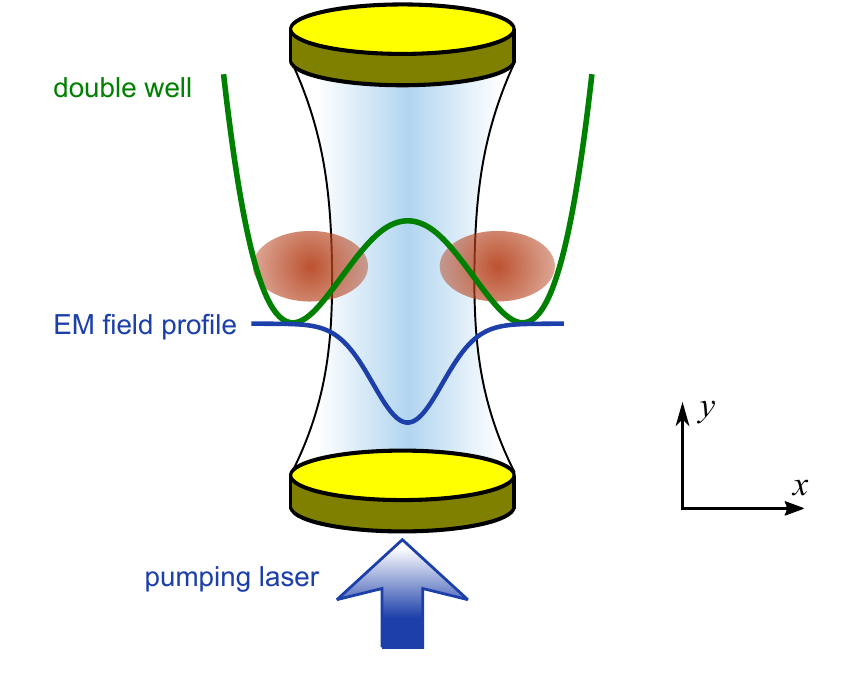}
\caption{(Color online) The illustration of the setup. 
The bosonic Josephson Junction is created by magnetic or optical means 
along the $x$ direction. A Fabry-P\'erot cavity is placed around the 
junction with an axis orthogonal to the junction. The resonator is 
operated on the TEM$_{00}$ mode.}
\label{fig:scheme}
\end{figure}

Here we consider a hybrid system consisting of a BJJ and an optical cavity, whose field is interacting with the atoms. The scheme is identical to that of Ref. \cite{articolo} and is illustrated in Fig.~\ref{fig:scheme}. The BJJ in our system is made of an interacting Bose-Einstein condensate in a symmetric double-well potential. Its Hamiltonian in second-quantized form reads as
\begin{equation}
\label{eq:Hamatom}
{\hat H}_A=\int d^3{\bf r}\ 
{\hat \Psi}^\dagger(\vec{r})\bigg[-\frac{\hbar^2}
{2m}\nabla^2+V(\vec{r})
+\frac{g}{2}{\hat \Psi}^\dagger(\vec{r}){\hat \Psi}(\vec{r})\bigg]{\hat \Psi}(\vec{r}),
\end{equation}
where ${\hat \Psi}(\vec{r})$ annihilates an atom at position $\vec{r}=(x,y,z)$. Low energy atom-atom scattering is characterized by $g=4\pi\hbar^2a/m$, where $a$ is the s-wave scattering length, and $m$ is the mass of an atom. The confining potential is
\begin{equation} 
\label{eq:pot}
V(\vec{r}) = V_{DW}(x) + {1\over 2} m \omega_{H}^2 (y^2+z^2) \; ,
\end{equation}
where $V_{DW}(x)$ is the double-well potential in the $x$ direction, while we assume tight harmonic confinement in the transverse direction, with trap frequency $\omega_{H}$.

The optical cavity is a single-mode high-Q Fabry-P\'erot resonator, whose axis is orthogonal to the double-well direction as depicted in Fig. \ref{fig:scheme}. We choose the $y$ direction for the cavity axis. The cavity has characteristic frequency $\omega_C$, it is pumped through one of its mirrors with a laser of frequency $\omega_L$ and amplitude $\eta$, and it is operated at the TEM$_{00}$ mode. The mode function of the cavity is
\begin{equation} 
f(\vec{r})= \sqrt{2\over L} \,\cos(k\,y) \, 
{e^{-(x^2+z^2)/(2\sigma^2)}\over \pi^{1/2} \sigma},
\end{equation} 
with $k=\omega_C/c$ is the wave number of the cavity mode, 
$L$ is the distance between the mirrors 
and $\sigma$ is the width of the Gaussian profile in the $(x,z)$ plane.

The cavity dynamics, in the frame rotating with the frequency $\omega_L$ of the laser drive, is governed by the Hamiltonian 
\begin{equation}
\hat H_{C}=- \hbar\Delta_C \hat N_L-i\hbar\eta(\hat a -\hat a^\dagger).
\end{equation}
Here $\Delta_C=\omega_L-\omega_C$ is the cavity detuning, $\eta$ is the pumping strength of the driving laser, $\hat a$ is the annihilation operator of a cavity photon and $\hat N_L=\hat a^\dagger\hat a$ is the photon number.

The condensate atoms have an electronic ground state and an excited state, with transition frequency $\omega_A$. We assume that the atomic detuning, $\Delta_A=\omega_L-\omega_A$, is large enough to the excited state population to be negligible all the time, and the atoms to behave as polarizable scalar particles. Therefore, their interaction with the cavity field is described by dispersive photon scattering \cite{ritch2013cold}, which shifts the cavity resonance frequency proportional to the atom number, and on the atomic degrees of freedom it creates an optical potential. This additional atom-light interaction term reads as
\begin{equation}
\label{eq:optpot}
\hat H_{AL}=\hbar U_0 \hat a^\dagger \hat a \int d^3{\bf r} f^2(\vec{r}){\hat \Psi}^\dagger(\vec{r}){\hat \Psi}(\vec{r}),
\end{equation}
where the strength of the interaction is $U_0=\Omega_R^2/\Delta_A$, with $\Omega_R$ the single-photon Rabi frequency.
As the cavity field is focused symmetrically to the center of the double-well barrier, the optical potential creates a light shift symmetric in the wells, and lowers (raises) the height of the barrier for red (blue) detuned atoms.

When the double-well barrier is sufficiently high, the lowest energy doublet is well separated from the rest of the spectrum. In this case the atomic dynamics can be constrained to the two-mode Fock space of the left and right valleys \cite{smerzi} and the atomic field operator can be approximated as
\begin{equation}
\label{eq:wannierexp}
{\hat \Psi}(\vec{r})= 
\left( w_1(x)\ {\hat b}_1+w_2(x)\ {\hat b}_2 \right) 
{e^{-(y^2+z^2)/(2l_H^2)} \over \pi^{1/2} l_H} \; ,
\end{equation}
where $\hat b_1$ and $\hat b_2$ are the annihilation operators of a boson in the Wannier-like states $w_{1}(x)$ and $w_{2}(x)$, centered around the minima of the left and right valleys, respectively. The 
characteristic length of the strong harmonic confinement in the $(y,z)$ plane is given by $l_H=\sqrt{\hbar/(m\omega_H)}$. In this limit, the Hamiltonian describing the BJJ dynamics is evaluated by inserting the field operator \eqref{eq:wannierexp} into the atomic Hamiltonian \eqref{eq:Hamatom},
\begin{equation}
\hat H_{\text{BJJ}}=\epsilon\, \hat  N - J \left( {\hat b}_1^{\dagger}{\hat b}_2 + {\hat b}_2^{\dagger} {\hat b}_1 \right)
+ {U\over 2} \left( \hat b_1^\dagger\hat b_1^\dagger\hat b_1\hat b_1+\hat b_2^\dagger\hat b_2^\dagger\hat b_2\hat b_2\right),
\end{equation}
The parameters $\epsilon$, $J>0$ and $U$ are the on-site energy of a single well, the tunneling amplitude and the on-site interaction energy, respectively. The values of these parameters are written in terms of overlap integrals of the Wannier functions \cite{articolo}.

Similarly, the atom light interaction \eqref{eq:optpot} evaluates to
\begin{equation}
\hat H_{I}=\hat N_L \left[ W_0 \hat N + W_{12} \left( {\hat b}_1^{\dagger} {\hat b}_2 + {\hat b}_2^{\dagger} {\hat b}_1 \right)\right].
\end{equation}
The term with $W_0$ is the symmetric AC-Stark shift, while the $W_{12}$ term is the consequence of the lowering of the barrier. This latter can be understood as a cavity assisted tunneling contribution.
The value of the parameters $W_0$ and $W_{12}$ are also expressed as overlap integrals \cite{articolo}. The full Hamiltonian of the system can be written as:
\begin{equation}
\begin{split}
\hat H &= \hat H_{\text{BJJ}} + \hat H_{C} + \hat H_{I}\\
&=(\epsilon+W_0 \hat N_L)\, \hat  N - (J-W_{12}\hat N_L) \left( {\hat b}_1^{\dagger}
{\hat b}_2 + {\hat b}_2^{\dagger} {\hat b}_1 \right) + \\
&+ {U\over 2} \left( \hat b_1^\dagger\hat b_1^\dagger\hat b_1\hat b_1+\hat b_2^\dagger\hat b_2^\dagger\hat b_2\hat b_2\right) - \hbar\Delta_C \hat N_L-i\hbar\eta(\hat a -\hat a^\dagger)\; ,
\end{split}
\label{eq:ham2q}
\end{equation}
A more detailed derivation of the Hamiltonian can be found in Ref. \cite{articolo}.
In this configuration, $W_0$ and $W_{12}$ can be either positive or negative since they are proportional to the atomic detuning $\Delta_A$ \cite{articolo}. If the atoms are red detuned, i.e. $\Delta_A<0$, both $W_0\,,W_{12}<0$; while for blue detuning, i.e. $\Delta_A>0$, one has $W_0\,,W_{12}>0$. The geometry of the system has a remarkable advantage: the cavity photons can influence the tunneling of the atoms between the two wells of the potential, as opposed to previous proposals of this type of system \cite{corney1998homodyne,zhang2008cavity,zuppardo2014cavity} which considered only the effect of the cavity photons on the on-site energies of the BJJ.

\section{Semiclassical approximation}
Let us assume that the system is in a full coherent state, in which both the atoms in the right and left well, and the cavity photons are described with coherent states: $|FCS\rangle = |\beta_1 \rangle \otimes |\beta_2\rangle \otimes |\alpha\rangle$, where ${\hat b}_j|\beta_j\rangle = \beta_j |\beta_j\rangle$ and $\hat{a}|\alpha\rangle =\alpha |\alpha\rangle$. It is convenient to write the eigenvalues of the atomic coherent state as: $\beta_j=\sqrt{N_j}\, e^{i\theta_j}$, where $N_j$ the average number of atoms in the $j$-th well and $\theta_j$ the corresponding phase, and similarly
$\alpha= \xi \, e^{i\phi}$, where $N_L=\xi^2$ is the average number of photons in the cavity and $\phi$ the corresponding phase. The dynamics of the system can be described by a set of four ordinary differential equations for the fractional imbalance of the atomic population of the wells $z=\frac{N_1-N_2}{N}$, the atomic relative phase $\theta=\theta_2-\theta_1$, and the photon variables $\xi$ and $\phi$ \cite{articolo}.
The photon dynamics can be adiabatically eliminated when $\delta_C=\Delta_C-N(W_0+W_{12}\sqrt{1-z^2}\cos\theta)/\hbar$ is orders of magnitude larger than $\hbar^{-1}\nu=(J-W_{12}\xi^2)/\hbar^2$ \cite{ritch2013cold}. When the magnitude of the cavity detuning $\Delta_C$ is much larger than that of $N W_0/\hbar$ and $N W_{12}/\hbar$ the system is described by:
\begin{subequations}
\label{eqs:odescost}
\begin{align}
\dot{z} &=-2\tilde\nu\sqrt{1-z^2}\sin\theta,\label{eq:odecostz}\\
\dot{\theta} &= \left(\tilde g + \frac{2\,\tilde\nu}{\sqrt{1-z^2}}\cos{\theta}\right)z,\label{eq:odecostheta}
\end{align}
\end{subequations}
where $\tilde \nu=(J-W_{12}{\xi}^2)/\hbar=(J-W_{12}\eta^2/\delta^2_C)/\hbar$.
When these conditions are satisfied, the photon field amplitude $\xi$ can be considered to be constant in time and $\xi=\eta/|\Delta_C|$. This set of equations is very similar to the Josephson equations for a bare Josephson junction. But here, because of the cavity photons, the tunneling amplitude $J$ is substituted with the assisted tunneling amplitude $\tilde J=J-W_{12}{\xi}^2$.
In the case of the bare BJJ the solutions to these equations can be divided into two classes. One is characterized by oscillations around $z=0$ and therefore the time average of the population imbalance is zero (Josephson regime). The second class of solutions is characterized by an average imbalance in the population of the wells (Self-trapping regime).
However, in our system, the cavity photons can be used to induce self trapping solutions for initial conditions and parameters that would not allow them in the bare junction.
When $U>0$ and $\tilde J>0$, the self trapping occurs when \cite{raghavan1999coherent}:
\begin{equation}
\label{selftrap}
\frac{{\Lambda}}{2}z^2(0)-\sqrt{1-z^2(0)}\cos\theta(0)>1.
\end{equation}
where $\Lambda=UN/2\tilde J$ and $z(0)$ and $\theta(0)$ are the initial conditions to the equations.
Since $\Lambda$ depends on the number of cavity photons $\xi^2$, in our system we have one more relevant parameter that allows us to switch between different regimes.
\begin{figure}
\centering
\includegraphics{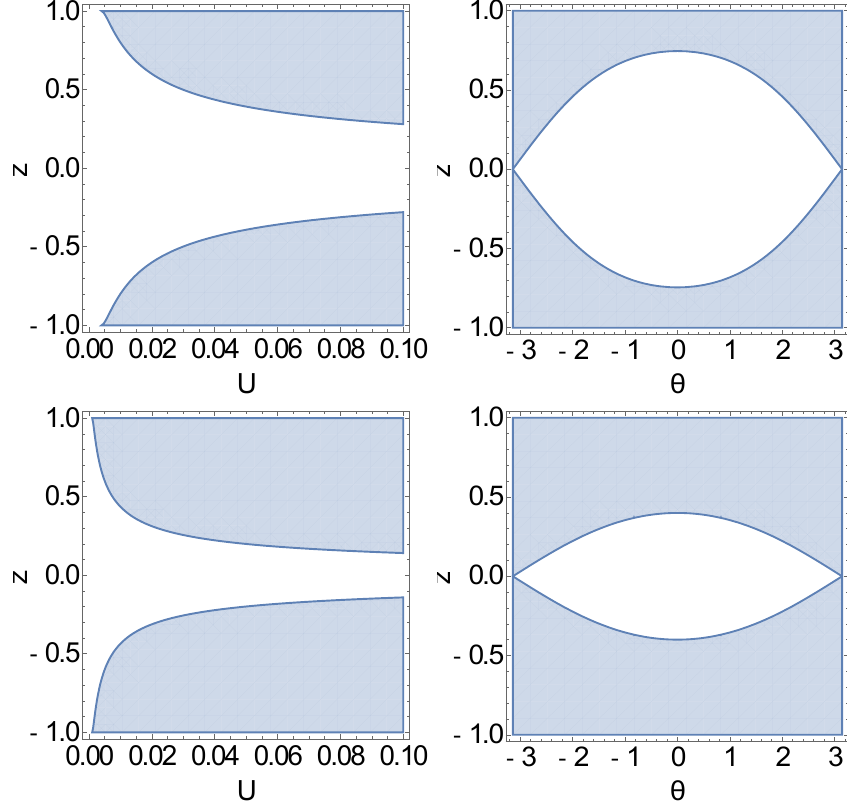}
\caption{\footnotesize (Color online) The highlighted area in the panels shows the values of $z$, $U$ and $\theta$ that allow self trapping. In the first column for a fixed $\theta=0$ and in the second column for a fixed $U=12J/N$. The difference between the two rows is the number of photons in the cavity. In the first one $\xi^2=0$ and $\tilde J=J$, in the second  $\xi^2=25$ and $\tilde J= 0.25J$. It is clear that, because of the cavity photons, the highlighted area gets wider and therefore the appearance of the self trapping is  assisted.  The other parameters are $N=1000$, $N W_{12}=30J$. $U$ is written in units of $J$.}
\label{blue}
\end{figure}
By comparing the panels in Fig. \ref{blue}, it can be clearly seen what the effect of changing the number of photons is when $W_{12}>0$. In particular we notice how increasing the number of photons widens the area that allows self trapping solutions.
\begin{figure}[!h]
\centering
\includegraphics{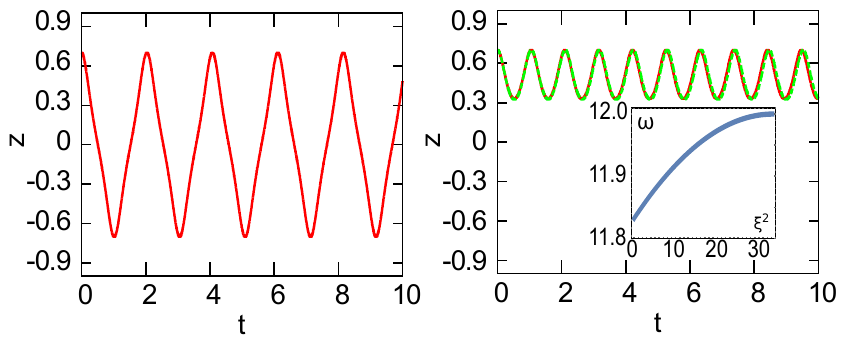}
\caption{\footnotesize (Color online) Time evolution of the variable $z$. Time is measured in units of $\hbar/J$.
The first panel represents the bare junction in the absence of the cavity. The second one shows the junction inside of the cavity.
The solid lines represent the numerical solution of the system (\ref{eqs:odescost}), where the photon variables are constant. The dashed line represents the solution of the system where the photon dynamics has not been adiabatically eliminated. When there are no photons, the system is in the Josephson regime, while when the photon number is increased the system is in the self trapped regime.  The parameters for the second panel are: $\hbar\Delta_C=300J$, $W_0 N=4 J$, $W_{12} N=3 J$, $U N=12 J $, $N=1000$, $\hbar\eta=3000 J$.
The initial conditions for the first panel are for $(z,\theta)=(0.7,0)$. The initial conditions for the second panel are $(z,\theta,\xi,\phi)=(0.7,0,10.17,\pi/2)$. The inset plot represents the frequency $\omega$ of the small oscillations of $z$ written as a function of the number of cavity photons $\xi^2$ . $\omega$ is written in units of $J/\hbar$. The range of $\xi^2$ is given by the condition for the system to show small oscillations. The other relevant parameters for the inset are $N = 1000$, $U N= 0.012J$ and  $W_{12} N = 0.03J$.}
\label{fig:induced}
\end{figure}
The photon-induced change of regime can be clearly seen from Fig. \ref{fig:induced}. The left panel shows the time evolution of $z$ in the Josephson regime, in the absence of photons. The right panel shows the time evolution when the cavity photons are present in the system: it can be clearly seen that the photons induce oscillations around a non-zero average value of $z$.
The cavity photons can be therefore used as a means of inducing self trapped solutions that would not otherwise occur in the absence of the cavity.

\section{Exact diagonalization}
We now want to show how the cavity photons influence the ground state of the system. In order to do this, we describe the photons with a coherent state $|\alpha \rangle$, but we describe the atoms in the wells in a purely quantum way. To this end we use Fock states as basis vectors in the atomic Hilbert space.
Furthermore, we assume that the photon field relaxes quickly and this allows us to consider the photon coherent state to be a parameter on which the atomic ground state depends.
The ground state of the atomic field for a given photon field coherent state $|\alpha \rangle$  can be written as: $|GS \rangle_{\alpha}= \sum_n c_n(\alpha)|n \rangle$,
where $|n \rangle=|N-n,n \rangle$ is an element of the Fock basis $\{|n\rangle, n=0,1,...,N\}$ and $N$ is the total number of atoms. The element of the Fock basis $|n\rangle$ describes the state in which $n$ atoms are in the right well and $N-n$ in the left well. The coefficients $c_n(\alpha)$ clearly depend on the photon field.
So the ground state of the system can be written as $|GS\rangle=|\alpha\rangle \otimes |GS \rangle_{\alpha}$, which is the tensor product of the coherent state of the photon field and of the atomic ground state for that field strength.
In order to calculate the ground state coefficients $c_n(\alpha)$
the matrix elements of the Hamiltonian that needs to be diagonalized are:
\begin{equation}
\small
\begin{split}
H_{m,n}(\alpha)=&-\left(J-W_{12}|\alpha |^2\right)(\sqrt{N-m}\sqrt{n}\,\delta_{m,n-1}\\
&+\sqrt{N-n}\sqrt{m}\,\delta_{m,n+1} )+\frac{U}{2}[(N-n)(N-n-1) \\
&+ n(n-1)]\delta_{m,n},
\end{split}
\label{matrix}
\end{equation}
If we assume that the photon field relaxes very fast, then $i\hbar \, {d\over dt} \alpha=0$ leads to a relation between $\alpha$ and the ground state coefficients $c_n(\alpha)$ that needs to be satisfied in order for the computation of the ground state to be consistent.
This makes the amplitude of the photon field $|\alpha\rangle$ dependent on the parameters that appear in the Hamiltonian and on the coefficients $c_n(\alpha)$ as well.
The analysis of the ground state can however be simplified when $|\Delta_C|\gg |W_0| N/\hbar$ and $|\Delta_C|\gg |W_{12}| N/\hbar$. When such conditions are satisfied $\alpha$ depends weakly on the coefficients $c_n(\alpha)$. Therefore, for any starting number of photons $|\alpha|^2$ the above procedure gives as result the starting $|\alpha|^2$.
The number of photons in the system can thus be modified by changing the values of $\eta$ and $\Delta_C$.
Under this approximation we only need to study how the ground states of the Hamiltonian with the assisted tunneling is influenced by the presence of the photons.
In the absence of the cavity, and thus $|\alpha|^2=0$ the ground state of the BJJ can show three limit behaviors depending on the value of the ratio $U/J$ \cite{gio5}. When $U/J=0$ the ground state is an atomic coherent state (ACS), where the distribution of the $|c_n(\alpha)|^2$ has a gaussian profile. For large values $|U/J|$ the ground state can be found in two configurations, depending on the sign of $U/J$. When $U/J>0$ the ground state is a separable Fock state, where half of the atoms are in the right well and half in the left one. When $U/J<0$ the ground state is a "cat state", which is the linear combination of the state with all particles in the left or in the right well. Changing the interaction strength $U/J$ leads to a continuous transitions between these three kinds of state.
\begin{figure}[t]
\centering
\includegraphics{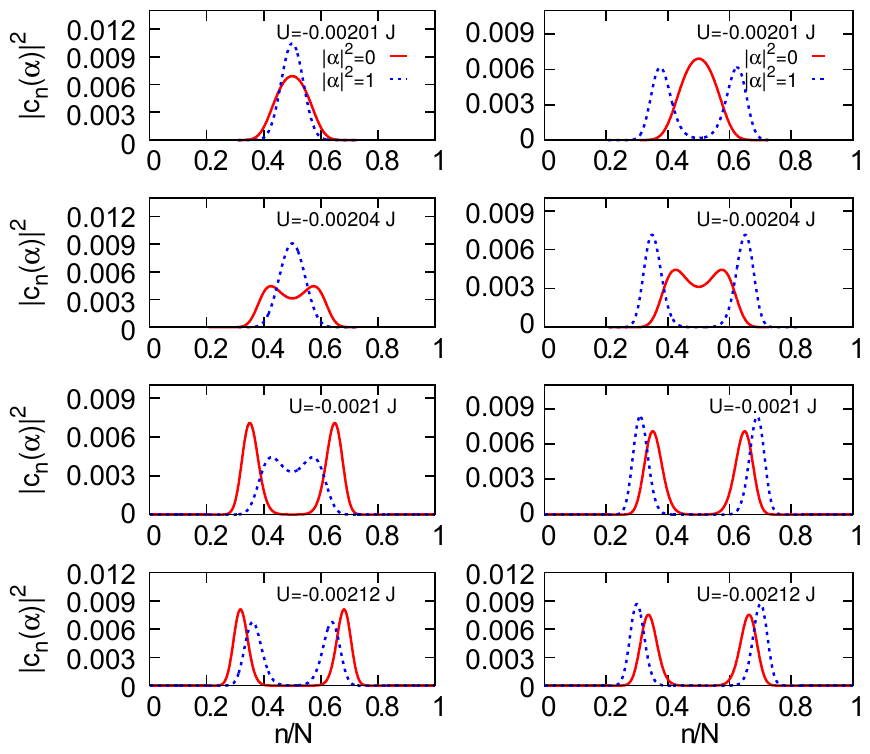}
\caption{\footnotesize (Color online) Coefficients $|c_n(\alpha)|^2$ of the ground state as a function of $n/N$ for $N=1000$. In panel (a) $W_{12}=-0.03J$ and in panel (b) $W_{12}=0.03J$. The on-site interaction is negative $U<0$. The solid line represents the system in the absence of cavity photons, $|\alpha|^2=0$, while the dashed line represents the system when $|\alpha|^2=1$.
The effect of the cavity photons in (a) is to increase the coherence of the ground state for a given negative on-site interaction $U$, and thus delay the transition to the regime in which a valley appears in the middle of the distribution of the coefficients $|c_n(\alpha)|^2$.
The effect of the cavity photons in (b) is the opposite. The coherence of the ground state decreases for a given negative on-site interaction $U$, and thus the photons assist the transition to the regime in which a valley appears in the middle of the distribution of the coefficients $|c_n(\alpha)|^2$.
In both cases the ground state is showed for four values of $U$. The coefficients  $|c_n(\alpha)|^2$ are adimensional and normalized so that $\sum_{n=0}^N |c_n(\alpha)|^2=1$. $n/N$ is adimensional.}
\label{fig:grafc}
\end{figure}
\begin{figure}[t]
 \centering
 \includegraphics{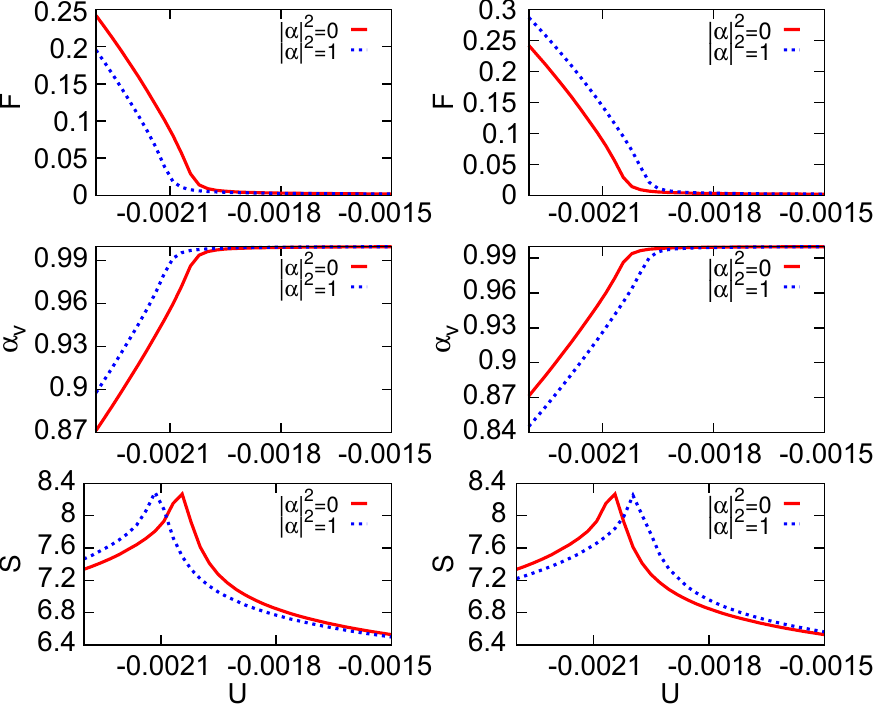}
  \caption{\footnotesize (Color online) Fisher information $F$, coherence visibility $\alpha_V$ and entanglement entropy S, plotted as a function of the on-site interaction $U$, which is written in units of $J$, for $N=1000$. In panel (a) $W_{12}=-0.03J$ and in panel (b) $W_{12}=0.03J$. The solid line represents the system when $|\alpha|^2=0$, while the dashed line represents the system when $|\alpha|^2=1$. The on-site interaction is attractive $U<0$. The plots of the quantum indicators complement those of Fig.  \ref{fig:grafc}. In the presence of photons $F$ is smaller when $W_{12}<0$ and larger when $W_{12}>0$, compared the the solid line, since $F$ is related to the distribution width of the $c_n(\alpha)$. $\alpha_v$ is larger for $W_{12}<0$ and smaller for $W_{12}>0$ compared to the solid line as the photons increase and decrease the coherence respectively. As for $S$, whose maximum represents the value of $U$ at which the distribution of the $|c_n(\alpha)|^2$ starts to show a valley \cite{gio5}, comes for a larger $|U|$ for $W_{12}<0$ and for a smaller $|U|$ for $W_{12}<0$. $F, \alpha_v$ and $S$ are dimensionless.}
  \label{fig:indicc}
\end{figure}

The correlation properties of the ground state can be analyzed by using the Fisher information $F$ \cite{pezze,weiss} and the coherence visibility $\alpha_v$ \cite{anna}. The definition for these two estimators can be found in \cite{gio5}. In particular, $F$ gives the width of the distribution of $|c_n(\alpha)|^2$ and $\alpha_v$ characterizes the degree of coherence between the two wells. We are interested in characterizing also the genuine quantum correlations between the atoms in two wells pertaining to the ground state. Here we focus on the atom part only and address this issue from the bi-partition perspective with the two wells playing the role of the two partitions. Following the same path as in \cite{gio5,galante} we thus calculate the entanglement entropy $S$ \cite{bwae} that results in $S=-\sum_{n=0}^{N}|c_{n}(\alpha)|^2\log_{2}|c_{n}(\alpha)|^2$. Remarkably, $S$, plotted as a function of $U$, shows a maximum at the onset of the transition to the "cat-like" state regime in which the distribution of the $|c_n(\alpha)|^2$ begins to exhibit a valley.
The Hamiltonian  we need to diagonalize in Eq. (\ref{matrix}) differs from the Hamiltonian of a Bosonic Josephson Junction only for the value of the tunneling amplitude, which we will call $\tilde J=J-W_{12}|\alpha |^2$. The magnitude and sign of $\tilde J$ depend on both $W_{12}$ and on the photon number $|\alpha|^2$.

When $W_{12}<0$ the assisted tunneling amplitude $\tilde J$ is always positive and its magnitude gets larger as  the number of photons in the system $|\alpha|^2$ increases.
Let us choose $W_{12}=-0.03 J$ and plot together the ground states for various values of $U$ for $|\alpha|^2=1$ and therefore $\tilde J=1.03 J$. We always plot the ground state of the bare junction, with $|\alpha|^2=0$, to use as a reference.
In panel (a) of Fig. \ref{fig:grafc} and of Fig. \ref{fig:indicc} we can see what happens when the on-site interaction $U$ is negative.
When the interaction is positive, the effect of the photons is still to delay the transition to the separable Fock state.

When $W_{12}>0$ the assisted tunneling amplitude $\tilde J$ gets smaller when increasing the number of photons in the cavity. This can lead to two main consequences. On the one hand, if the number of photons is sufficiently small, $\tilde J$ maintains its positive sign but its magnitude gets smaller  and $0<\tilde J/J<1$. On the other hand if the number of photons is large enough the assisted tunneling can become negative $\tilde J/J<0$.
When $0<\tilde J/J<1$ we can plot the ground states of the system, for different values of negative $U$, and for $|\alpha|^2=1$, and therefore $\tilde J=0.97 J$. We can see from panel (b) of Fig. \ref{fig:grafc} and of Fig. \ref{fig:indicc} that the effect of the photons in this case is opposite to the effect when $W_{12}$ is negative.
When the interaction is positive, the effect of the photons is still to assist the transition to the separable Fock state.
It is worth mentioning that this result is actually more interesting then the sole rescaling of the ratio $U/J$. Focusing our attention on the attractive on-site interaction and on the emergence of the "cat-like" state regime we can notice a fundamental difference. For a fixed number of atoms in the system, in the bare Josephson junction, the transition to the "cat-like" state regime happens for a definite value of $U/J$. However, in the complete system, with a positive $W_{12}$, by fine tuning the number of cavity photons the transition to the "cat-like" state regime can happen for a given magnitude of negative $U$, without changing the number of atoms.
This means that the photons in the system act as a new knob through which the transition between different regimes can be manipulated.
The change of the sign of $\tilde J$ is significant when analyzing the Hamiltonian of the system. In the case of the bare Josephson junction $J$ can always be taken to be positive.
However, by numerically calculating the ground state of the system we found out that the sign of $\tilde J$ bears no relevance to the coefficients $|c_n(\alpha)|^2$ and to the value of the quantum indicators. Therefore, choosing values of $|\alpha^2|$ such that the values of the assisted tunneling amplitude $\tilde J$ is the same as the ones we have just studied, but with opposite signs, leads to the same results.
It is interesting to notice this analogy between the case with positive $\tilde J$ and negative $\tilde J$. One would at first think that, since the relevant parameter in the analysis is the ratio $U/\tilde J$, when $\tilde J$ is negative, the same results could be obtained by changing the sign of $U$ as well, however this is not the case.

\section{Summary}
To summarize, we have shown that the cavity photons can be used to induce self-trapping solutions in the dynamics of the Josephson junction. Moreover we have described how the transition from the atomic coherent state to a "cat state" can be delayed or expedited, depending on the photon-assisted tunneling amplitude. This analysis lays the first stone of the quantum study of this system. Possible extensions to this work could be the study of the ground state when the $|\Delta_C|\gg |W_0| N/\hbar$ and $|\Delta_C|\gg |W_{12}| N/\hbar$ conditions no longer hold. This would lead to the same ground states we have already found, but the photon number of the cavity would depend on the other parameters of the system.
Moreover one could try to better understand the transition between the atomic coherent state and the cat state, by computing the relevant critical indices for $F$ and $\alpha_v$.

The setup considered here can be implemented directly due to the current feasibility of trapping ultracold bosons in double-well potentials \cite{albiez2005direct,schumm2005matter,levy2007ac} and to fabricate high-finesse optical cavities supporting quantum degenerate bosonic clouds \cite{brennecke2007cavity,colombe2007strong,murch2008observation}. The experiment deriving from the suitable combination of these two parts would provide a powerful tool for observing the Josephson--self-trapping crossover, an issue addressed with bare BJJs \cite{albiez2005direct} and creating entangled states of matter. In this context, experimentalists could have the concrete possibility to receive feedback on the shift of the crossover above by using the number of cavity photons as a knob. The setup considered in Ref. \cite{colombe2007strong} is extremely appealing, where the Fabry-P\'erot cavity is assembled on top of an atom chip, where wires suitable for the double-well potential can be easily placed too.
On such a device one has to fine tune experimental parameters that mainly consist in the proper matching of the double-well dimensions (a barrier width of a few microns) with those of the waist of the cavity (also in the range of a few microns).

\section*{Acknowledgements}
The authors acknowledge Italian Ministry of Education,
University and Research (MIUR) for partial support (PRIN Project 2010LLKJBX
"Collective Quantum Phenomena:  from Strongly-Correlated Systems to Quantum Simulators").
G. Sz. also acknowledges support from the Hungarian Scientific Research Fund (Grant No. PD104652), and the János Bolyai Scholarship.

\end{document}